\documentclass[conference]{IEEEtran}
\IEEEoverridecommandlockouts

\usepackage{cite}
\usepackage{float}
\usepackage{amsmath,amssymb,amsfonts}
\usepackage{algorithmic}
\usepackage{graphicx}
\usepackage{textcomp}
\usepackage{xcolor}
\usepackage{soul}
\usepackage{booktabs}
\usepackage{multirow}

\def\BibTeX{{\rm B\kern-.05em{\sc i\kern-.025em b}\kern-.08em
    T\kern-.1667em\lower.7ex\hbox{E}\kern-.125emX}}

\begin{document}

\title{PiRL: Participant-Invariant Representation Learning for Healthcare Using Maximum Mean Discrepancy and Triplet Loss \\

\thanks{This work is supported by NSF \#2047296 and \#1840167.}
}

\author{\IEEEauthorblockN{Zhaoyang Cao}
\IEEEauthorblockA{\textit{Department of Computational and Applied Mathematics} \\
\textit{Rice University}\\
Houston, US. \\
zc48@rice.edu}
\and
\IEEEauthorblockN{Han Yu}
\IEEEauthorblockA{\textit{Department of Electrical and Computer Engineering} \\
\textit{Rice University}\\
Houston, US. \\
hy29@rice.edu}
\and
\IEEEauthorblockN{Huiyuan Yang}
\IEEEauthorblockA{\textit{Department of Electrical and Computer Engineering} \\
\textit{Rice University}\\
Houston, US. \\
hy49@rice.edu}
\and
\IEEEauthorblockN{Akane Sano}
\IEEEauthorblockA{\textit{Department of Electrical and Computer Engineering} \\
\textit{Rice University}\\
Houston, US. \\
akane.sano@rice.edu}

}

\maketitle

\begin{abstract}
Due to individual heterogeneity, person-specific models are usually achieving better performance than  generic (one-size-fits-all) models in data-driven health applications. 
However, generic models are usually preferable in real-world applications,  due to the difficulties of developing person-specific models, such as new-user-adaptation issues and system complexities. 
To improve the performance of generic models, we propose a \textbf{P}articipant-\textbf{i}nvariant \textbf{R}epresentation \textbf{L}earning (PiRL) framework, which utilizes maximum mean discrepancy (MMD) loss and domain-adversarial training to encourage the model to learn participant-invariant representations. 
Further, to avoid trivial solutions in the learned representations, a triplet loss based constraint is used for the model to learn the label-distinguishable embeddings.
The proposed framework is evaluated on two public datasets (CLAS and Apnea-ECG), and significant performance improvements are achieved compared to the baseline models. 
\end{abstract}


\section{Introduction}

Deep learning has gained popularity in modeling time-series data for solving health-related problems.  
For example, Oh \textit{et al.}\cite{oh2018automated} proposed an automated system that combines a convolutional neural network (CNN) and a long short-term memory network (LSTM) for the diagnosis of arrhythmia. Erdenebayar \textit{et al.}\cite{erdenebayar2019deep} designed a deep neural network, recurrent neural networks, and a gated-recurrent unit to distinguish apnea and hypopnea events using an electrocardiogram (ECG) signal.
In addition to physical health, deep models have also been used for mental health.  Yu and Sano \cite{yu2022semi} applied semi-supervised learning on leveraging unlabeled data to estimate wearable-based momentary stress. Radhika \textit{et al.}\cite{radhika2020transfer,radhika2021deep} proposed the frameworks that investigate the effectiveness of transfer learning and deep multimodal fusion on CNN stress detection models.

Although deep learning models have achieved promising results in health applications, researchers have observed that  person-specific models usually outperform generic models \cite{gjoreski2016continuous, kogler2015psychosocial, nakashima2015stress, nkurikiyeyezu2019effect, schmidt2018introducing, zenonos2016healthyoffice}, 
due to heterogeneous time series data. 
For example, Bsoul \textit{et al.}\cite{bsoul2010apnea} showed that the accuracy of the subject-dependent sleep apnea classification model is 6\% higher than that of the subject-independent model.
Nath \textit{et al.}\cite{nath2020comparative} showed a performance gap of 22.5\% in accuracy between the subject-dependent and the generic LSTM models in emotion recognition. 
Although person-specific models have been widely proven to outperform the generic models in health applications \cite{healey2005detecting, koldijk2016detecting,valenza2014revealing}, we cannot therefore ignore the challenges of developing person-specific models. For example, person-specific models cannot be easily extended to unseen participants\cite{nkurikiyeyezu2019effect}, and it is usually impractical to collect enormous datasets from individuals to build person-specific models.

Researchers have explored improving the performance of generic models by introducing person-specific information.
For example, Radhika \textit{et al.}\cite{radhika2020transfer, radhika2021deep} used person-specific information in the testing set during the feature extraction and selection. Wu \textit{et al.}\cite{wu2018personalizing} achieved model personalization from a pre-trained generic model by active learning approach and improved detection precision on
each new patient. Similarly, 
Li \textit{et al.}
\cite{li2018patient} fine-tuned a generic convolutional neural network to the tuned dedicated CNN for 
extracting the  characteristic information of a specific patient and obtained the effectiveness on arrhythmia prediction.
Bethge \textit{et al.} \cite{bethge2022domain} utilized MMD loss to impose domain-invariant representations for emotion classification tasks, where each participant had his/her own private encoder with a classifier shared among all. 
Although improved performances  were observed in the generic models, problems still exist in those studies, as individual encoders result in both high computational costs and difficulties in adapting to new subjects.

In this work, we aim to improve the performances of generic models without introducing extra computational complexity into the model, thus avoiding the aforementioned drawbacks of person-specific methods. We propose a representation learning framework to learn participant-invariant features through two learning phases, phase I: unsupervised learning and phase II: supervised learning. During the unsupervised learning phase, maximum mean discrepancy (MMD) loss is integrated with representation learning, which alleviates the heterogeneous issues and the influence of person-specific information, to minimize the distribution shifts among features from different subjects. Instead of MMD loss, we also test domain classification loss from domain-adversarial training of neural networks (DANN) architecture, which aims to blur the participant-distinguishable information among the learned representations, to make label predictor more robust to the target participant \cite{bethge2022domain}. During the supervised learning phase, triplet loss, which aims to learn the label-distinguishable embedding, is also used in the framework as a constraint to avoid trivial solutions in learned representations. We evaluate the proposed framework using two public datasets, including downstream tasks: sleep apnea detection and stress detection. Our results suggest that the proposed method can help improve the model performance significantly compared to the baseline representation learning model.  


\section{Methodology}
\begin{figure*}[!htbp]
\centering\includegraphics[width=0.85\textwidth,height=6.2cm]{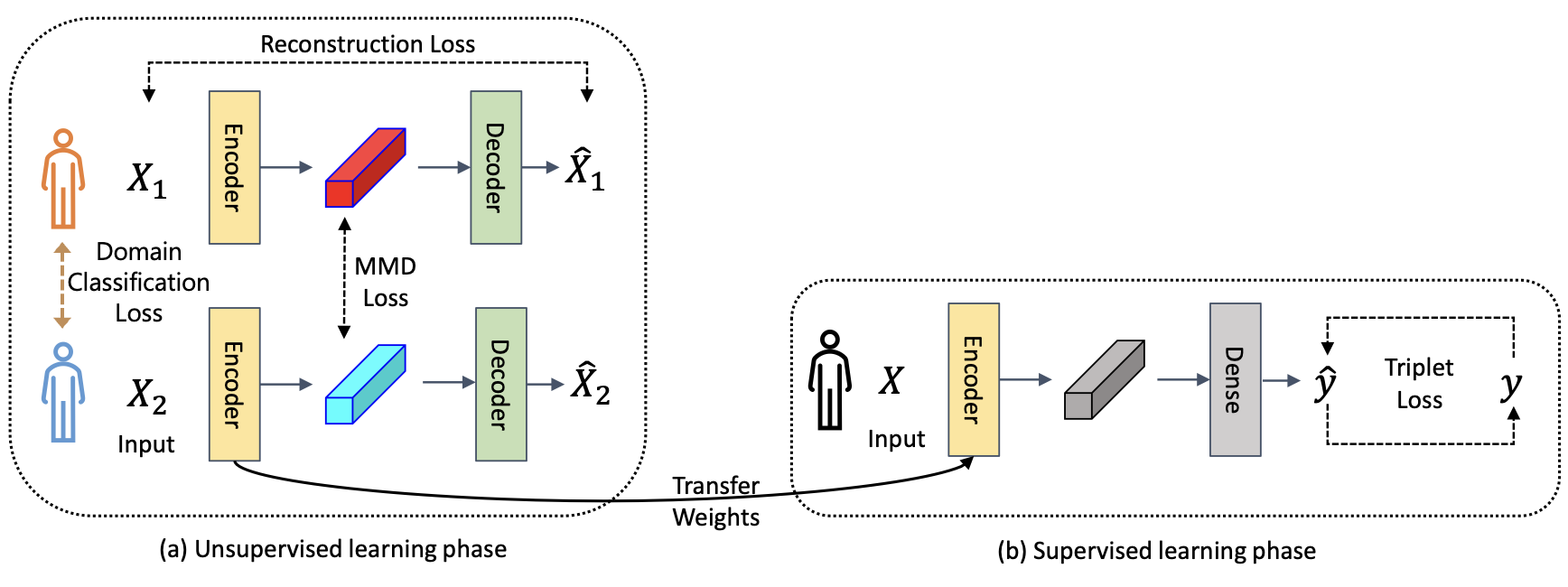}
    \caption{\textbf{PiRL Network Architecture} (a) step I: unsupervised learning which consists of a 1D CNN-based auto-encoder with reconstruction loss and penalized MMD loss and domain classification loss. (b) step II: supervised learning which utilized triplet loss for label-distinguishable
    representation.}
    \label{fig:PiRL}
\end{figure*}

We propose a representation learning framework that aims to extract participant-invariant representations, named PiRL. The main PiRL framework is visualized in Figure \ref{fig:PiRL} and detailed architecture is shown in Experimental Setting \ref{model_description}. We employ a 1D CNN-based auto-encoder structure as a deep feature extractor from input wearable data. On top of the auto-encoder, MMD loss and domain classification loss are utilized to constrain the representations from distribution shifts, therefore encouraging the learning embedding to be participant-invariant. During training for downstream tasks, we optimize the model with a triplet loss for label-distinguishable representations. The following subsections will introduce the aforementioned components in detail. 

\subsection{Representation learning}
We utilize a 1D CNN-based auto-encoder to extract learning representations from raw time-series sequences $X$. The encoder extracts latent representations from input sequences as $e$ with a series of 1D CNN layers; whereas the decoder aims to output the reconstructed signal as $\hat{X}$ from $e$ using up-sampling layers. The objective of the auto-encoder is:
\begin{equation}
    \mathcal{L}_{ae} = \|X - \hat{X}\|_2^2
\end{equation}

\subsubsection{MMD loss} 
To encourage the model to learn the participant-invariant representations, we constrain the model with an MMD loss function, which is widely used in eliminating distribution shifts among different groups of data \cite{gretton2012kernel}.

Given training samples from two different subjects as $X_i$ and $X_j$ with the total number of subjects $N$, the MMD loss can be considered as follows:
\begin{center}
$\mathcal{L}_{mmd}(p,q,\mathcal{H}) = \sum_{i=1}^{N}\sum_{j=1}^{N}\sup\limits_{f \in \mathcal{H},||f||_{\mathcal{H}}\leq1}(
\mathop{\mathbb{E}}\limits_{p(X_i)}[f(X_i)] -\mathop{\mathbb{E}}\limits_{q(X_j)}[f(X_j)])$
\end{center}
where $\textit{p}$ and $\textit{q}$ are the distributions of variable $X_i$ and $X_j$, $\mathcal{H}$ is the Hilbert space, and $\textit{f}$ is the mapping function. During the training process, the model is optimized to minimize the distribution distances between the pairs of subjects. Thus, the overall objective of unsupervised learning is:
\begin{equation}
    \mathcal{L} = \mathcal{L}_{ae} + \lambda \cdot \mathcal{L}_{mmd}
\end{equation}
where $\lambda$ is the coefficient of MMD loss and set to 0.2.

 \subsubsection{Domain classification loss} 

A domain classifier in DANN \cite{bethge2022domain} can help learn a classifier that is more robust in target data by 
accomplishing the following adversarial tasks: minimizing the loss of label prediction while maximizing the loss of domain classification. 
Subsequently, with the addition of a gradient reversal layer before the domain classifier, the overall objective function is the sum of two minimizing problems \cite{ganin2016domain}. 
In our case, we treat the individual participant as an independent domain, and the number of domains is equal to the number of participants. The corresponding domain classification loss is given below:
\begin{equation}
    \mathcal{L}_{domain} = -\sum_{i=1}^{N}(p_i\cdot log(q_i))
\end{equation}
where $\textit{N}$ is the number of labels, $p_i$ is the true probability distribution of one-hot encoding of the source domain, and $q_i$ is the predicted probability distribution of one-hot encoding computation of the target domain. As a result, the overall objective of unsupervised learning is:
\begin{equation}
    \mathcal{L} = \mathcal{L}_{ae} + \mu \cdot \mathcal{L}_{domain}
\end{equation}
where $\mu$ is the coefficient of domain classification loss and set to -1. At last, the final objective function of this study is \begin{equation}
    \mathcal{L} = \mathcal{L}_{ae} +  \lambda \cdot \mathcal{L}_{mmd}+ \mu \cdot \mathcal{L}_{domain}
\end{equation}

\subsection{Fine-tuning with triplet loss}
The supervised learning models for downstream tasks are built on top of the learned representations and in this work, a fully connected layer is used as a classifier.
Then the supervised learning structure is fine-tuned according to different downstream tasks. Furthermore, we also apply the triplet loss  \cite{schroff2015facenet} to optimize the representation of the training labels and avoid the trivial solutions learned in the pre-training procedure. The triplet loss is given below:
\begin{equation}
\label{eq:triplet}
\mathcal{L}_{triplet} = \max (d(a,p)-d(a,n)+margin, 0)
\end{equation}
where $\textit{a}$ represents anchor sample data, 
$\textit{p}$ represents `positive' sample data with the same class label from the anchor, $\textit{n}$ represents `negative' sample data with different class labels against the anchor, and the margin is a positive scalar. From the Eq-\ref{eq:triplet}, we can see that the objective of the triplet loss is to pull samples with the same class labels closer together in embedding space while pushing dissimilar ones apart.

\subsection{Latent Space Similarity}
Similarity of participants' embeddings in the latent space is often helpful in demonstrating the relationships among participants and how the representations perform in
a reduced-dimensional space. Since cosine and manhattan distance are two commonly used approaches to depict and quantify the latent space similarities, we used these two types of distance measures to evaluate the effects of MMD loss. We evaluated the pairwise distance of participants' representations in the test set to measure the latent space similarities. 
The formula of the cosine and manhattan distance is given as follows:
\begin{equation}
    D_{Cosine}(x,y) = 1- \frac{<x,y>}{||x||_2\cdot||y||_2}
\end{equation}
\begin{equation}
    D_{Manhattan}(x,y) = \sum_{i=1}^{n}|x_i - y_i|
\end{equation}
where \textit{x} and \textit{y} are both given vectors with the same dimension. In order to obtain the pairwise distance, we first used t-distributed stochastic neighbor embedding (t-SNE) to transform the representations into a 2-dimensional vector in this study as it is commonly used to realize the dimension reduction for high-dimensional data \cite{van2008visualizing}. As cosine and manhattan distance requires input vectors to be in the same dimensionality, we flattened the latent space of each participant from the baseline and MMD models to the 1 $\times$ 32 vector space which is used for the computation.

\section{Experimental  Setting}\label{experimental_setting}
Two public datasets, including CLAS \cite{markova2019clas} (stress detection) and Apnea-ECG \cite{penzel2000apnea} (sleep apnea detection), are used to evaluate the proposed framework.

\subsection{Dataset}
\label{dataset_info}
\subsubsection{CLAS Dataset}
The CLAS \cite{markova2019clas} dataset was collected for the automatic assessment of certain states of mind and emotional conditions using physiological data.
The dataset consists of recordings of electrocardiogram (ECG), photopletysmogram (PPG), electrodermal activity (EDA), and acceleration (ACC) signals.
The dataset was collected from 62 healthy subjects who participated in three interactive tasks and two perceptive tasks.
Labels for arousal, valence, and stress were assigned based on the stimuli tags for the interactive tasks.
In this paper, we conducted experiments for stress detection (class 0: non-stressed, class 1: stressed) using EDA signals.

The raw EDA data were first prepossessed with a low pass Butterworth filter (a cutoff frequency of 0.2 Hz), and then split into 10-second segments. The train/test set was split in a subject-independent manner.
Table \ref{table: datasetpreparationclas} summarizes the EDA training and testing sets used for the experiments. 

\begin{table}[!htbp]
\centering
\caption{Details of the training and the testing set in the CLAS dataset}
\label{table: datasetpreparationclas}
\begin{tabular}{l|ccc}
\toprule
    & \#Participants & \#Non-stressed & \#Stressed\\
    \midrule
    Training Set & 45 & 746 & 247\\
     \midrule
     Testing Set& 13 & 269 & 90\\
     \midrule
     Total& 58 & 1015  & 337 \\
     \bottomrule
\end{tabular}
\end{table}

\subsubsection{Apnea-ECG Dataset}
The recordings of Apnea-ECG \cite{penzel2000apnea} dataset include continuous ECG signals and sets of annotations for apnea (respiratory signals) from 70 participants. 
The recordings of 35 participants were used as a training set, and the rest was used as a testing set.
The length of the recordings ranged from slightly less than 7 hours to about 10 hours each, and the labels were provided as indicators of the presence (class 1) or absence (class 0) of sleep apnea in each minute of each recording. Thus, we split the ECG recordings into individual one-minute segments.
Table \ref{table: datasetpreparationecg} summarizes the details about the training and the test sets used in our experiment. 

\begin{table}[!htbp]
\centering
\caption{Details of the training and the testing set in the Apnea-ECG dataset}
\label{table: datasetpreparationecg}
\begin{tabular}{l|ccc}
\toprule
  & \#Participants & \#Normal breath& \#Disordered
breath \\
    \midrule
    Training Set & 35 & 10496 & 6514\\
     \midrule
     Testing Set& 35 & 10685 & 6548\\
     \midrule
     Total& 70 & 21181  & 13062 \\
     \bottomrule
\end{tabular}
\end{table}

For the CLAS and Apnea-ECG datasets, we used instance-wise min-max normalization for each time series segment on the training set and testing set separately and independently to ensure that each normalized segment had a similar scale within (0,1). 

\subsection{Model Description}
\label{model_description}
\subsubsection{Step 1: Unsupervised Learning}
For the CLAS dataset, the encoder
consists of 9 CNN layers and a fully connected layer (named 'Embedding'), whereas the decoder consists of corresponding 8 up-sampling layers.
For the Apnea-ECG dataset, the encoder consists of 12 CNN layers and an Embedding layer, whereas the decoder consists of corresponding 11 up-sampling layers. The detailed structures of the framework including the shape of input/output, kernel size, normalization method, and activation function are shown in Table \ref{table: structure_clas} and \ref{table: structure_apneaecg} for the CLAS and Apnea-ECG dataset respectively. 


\begin{table}[!htbp]
\centering
\caption{ Structure of the framework for CLAS dataset. B: batch size; L: length
of sequence}
\label{table: structure_clas}
\begin{tabular}{cccc}
\toprule
  Layer Name & Input Shape & Output Shape& Kernel \\
    \midrule
    Conv1d-1 & [B, 1, L] & [B, 32, L/2] & [1x4]\\
    \midrule\multicolumn{4}{c}{BatchNorm, ReLU}\\
     \midrule
     Conv1d-2& [B, 32, L/2] & [B, 64, L/4] & [1x4]\\
     \hline\multicolumn{4}{c}{BatchNorm, ReLU}\\
     \midrule
     Conv1d-3& [B, 64, L/4] & [B, 64, L/8] & [1x4]\\
     \hline\multicolumn{4}{c}{BatchNorm, ReLU}\\
     \midrule
     Conv1d-4& [B, 64, L/8] & [B, 128, L/16]  & [1x4] \\
     \midrule\multicolumn{4}{c}{BatchNorm, ReLU}\\
      \midrule
     Conv1d-5 & [B, 128, L/16] & [B, 128, L/32]  & [1x4] \\
     \midrule\multicolumn{4}{c}{BatchNorm, ReLU}\\
     \midrule
     Conv1d-6 & [B, 128,L/32] & [B, 256, L/64]  & [1x4]\\
     \midrule\multicolumn{4}{c}{BatchNorm, ReLU}\\
     \midrule
     Conv1d-7 & [B, 256, L/64] & [B, 256, L/128]  & [1x4]\\
     \midrule\multicolumn{4}{c}{BatchNorm, ReLU}\\
     \midrule
     Conv1d-8 & [B, 256, L/128] & [B, 512, L/256]  & [1x4]\\
     \midrule\multicolumn{4}{c}{BatchNorm, ReLU}\\
     \midrule
     Conv1d-9 & [B, 512, L/256] & [B, 512, L/512]  & [1x4]\\
     \midrule\multicolumn{4}{c}{BatchNorm, ReLU}\\
     \midrule
     Embedding (fc) & [B, 512] & [B, 8]  &  -\\
     \midrule
     fc-1 & [B, 512] & [B, 32]  & - \\
     \midrule\multicolumn{4}{c}{ReLU}\\
     \midrule
     fc-2 & [B, 32] & [B, 2]  & - \\
     \bottomrule
\end{tabular}
\end{table}

\begin{table}[!htbp]
\centering
\caption{ Structure of the framework for Apnea-ECG dataset. B: batch size; L: length
of sequence}
\label{table: structure_apneaecg}
\begin{tabular}{cccc}
\toprule
  Layer Name & Input Shape & Output Shape& Kernel \\
    \midrule
    Conv1d-1 & [B, 1, L] & [B, 16, L/2] & [1x4]\\
    \midrule\multicolumn{4}{c}{BatchNorm, ReLU}\\
     \midrule
     Conv1d-2& [B, 16, L/2] & [B, 32, L/4] & [1x4]\\
     \midrule\multicolumn{4}{c}{BatchNorm, ReLU}\\
     \midrule
     Conv1d-3& [B, 32, L/4] & [B, 64, L/8] & [1x4]\\
     \midrule\multicolumn{4}{c}{BatchNorm, ReLU}\\
     \midrule
     Conv1d-4& [B, 64, L/8] & [B, 64, L/16]  & [1x4]\\
     \midrule\multicolumn{4}{c}{BatchNorm, ReLU}\\
     \midrule
     Conv1d-5 & [B, 64, L/16] & [B, 128, L/32]  & [1x4] \\
     \midrule\multicolumn{4}{c}{BatchNorm, ReLU}\\
     \midrule
     Conv1d-6 & [B, 128, L/32] & [B, 128, L/64]  & [1x4] \\
     \midrule\multicolumn{4}{c}{BatchNorm, ReLU}\\
     \midrule
     Conv1d-7 & [B, 128, L/64] & [B, 256, L/128]  & [1x4] \\
     \midrule\multicolumn{4}{c}{BatchNorm, ReLU}\\
     \midrule
     Conv1d-8 & [B, 256, L/128] & [B, 256, L/256]  & [1x4] \\
     \midrule\multicolumn{4}{c}{BatchNorm, ReLU}\\
     \midrule
     Conv1d-9 & [B, 256, L/256] & [B, 512, L/512]  & [1x4] \\
     \midrule\multicolumn{4}{c}{BatchNorm, ReLU}\\
     \midrule
     Conv1d-10 & [B, 512, L/512] & [B, 512, L/1024]  & [1x4]\\
     \midrule\multicolumn{4}{c}{BatchNorm, ReLU}\\
     \midrule
     Conv1d-11 & [B, 512, L/1024] & [B, 1024, L/2048]  & [1x4]\\
     \midrule\multicolumn{4}{c}{BatchNorm, ReLU}\\
     \midrule
     Conv1d-12 & [B, 1024, L/2048] & [B, 1024, L/4096]  & [1x4]\\
     \midrule\multicolumn{4}{c}{BatchNorm, ReLU}\\
     \midrule
     Embedding (fc) & [B, 1024] & [B, 8]  & - \\
     \midrule
     fc-1 & [B, 1024] & [B, 32]  & - \\
     \midrule\multicolumn{4}{c}{ReLU}\\
     \midrule
     fc-2 & [B, 32] & [B, 2]  & - \\
     \bottomrule
\end{tabular}
\end{table}

\subsubsection{Step 2: Supervised Learning}
The same encoder structure was used in supervised learning (step2), but with two extra fully connected layers (fc-1 and fc-2 shown in Table \ref{table: structure_clas} and \ref{table: structure_apneaecg}) as a classifier.

\subsubsection{Person-specific Models}

The performance of the person-specific models was calculated to compare against the baseline and the proposed PiRL frameworks. We used the original training set of CLAS dataset to obtain the performance of person-specific models. Specifically, as shown in Table \ref{table: datasetpreparationspecific}, for each participant, the original training set was divided into a new training set and testing set with a ratio of 70\%$:$30\%. The person-specific models were trained on the training sets (Tr1) and tested on the testing sets (Te1) to obtain the results. The final prediction accuracy of the person-specific models was reported as the average and standard deviation of accuracy of all participants. 
\begin{table*}[!htbp]
\centering
\caption{Person-specific models data separation}
\label{table: datasetpreparationspecific}
\begin{tabular}{l|ccc}
\toprule
 &  \#Samples in EDA dataset & \#Samples in Apnea-ECG dataset\\
    \midrule
    Training Set (Tr1) & 695 & 11907\\
     \midrule
     Testing Set (Te1)& 298 & 5103\\
     \midrule
     Total& 993 & 17010 \\
     \bottomrule
\end{tabular}
\end{table*}



\subsection{Training Process}


In step 1, the unsupervised learning phase, we used an Adam optimizer with an initial learning rate of 0.001, and the learning rate was decayed by 0.9 after every 5 epochs.
The batch size was 32 for the CLAS dataset and 256 for the Apnea-ECG dataset. The length of sequence for the CLAS dataset was 960 and 6000 for the Apnea-ECG dataset. We trained the model for 100 epochs. The optimal weight for MMD loss was selected to be 0.2 from 0.1 to 0.5 with an increase of 0.1 in cross validation.

In step 2, the supervised learning phase, 
 the parameters such as batch size and learning rate remained the same, and four supervised models: 1) baseline model, 2) MMD loss only, 3) triplet loss only, 4) MMD and triplet loss, were evaluated. 
 The optimal weight for the triplet loss was selected to be 0.2 from 0.1 to 0.5 with an increase of 0.1 in cross validation.
 To evaluate the performance,  accuracies and standard deviations of 10 runs were reported.

Our model was implemented in the Pytorch deep learning framework, and was trained and tested on the NVIDIA GeGorce 3090Ti GPU.

\section{Results and Discussion}
We tested the proposed PiRL frameworks on two datasets, including CLAS and Apnea-ECG datasets for applications in mental health and physical health.
Experimental Setting \ref{experimental_setting} already included detailed information on two datasets and experimental settings such as hyper-parameters. For each supervised prediction model, we pre-trained and fine-tuned the encoder at the beginning of each epoch. We compared the prediction accuracies of the PiRL models against the ones of the baseline (only an auto-encoder and a supervised learning model without any additional constraints) and person-specific models.

\subsection{Stress Detection using Electrodermal
Activity (EDA) with CLAS Dataset}
Table \ref{table: clas} shows the prediction accuracy in all types of supervised learning models. To examine the statistical significance of the accuracy, we conducted an ANOVA (post-hoc: Tukey) test, and the corresponding results are also included in Table \ref{table: clas}. The prediction accuracy of the baseline framework was treated as the reference group.
The domain classification loss-based model did not show a significant increase in accuracy. The MMD loss-based model showed a higher accuracy of 66.5\% compared to the baseline results (64.3\%). Additionally, the prediction accuracy of the triplet loss only and the MMD + triplet loss models both exceeded 70\%.  The results illustrated that the model with triplet loss works better and improved the model performance most obviously in stress prediction than the one with MMD loss-based models. As expected, the person-specific models showed the highest accuracy (86.8\%) but also the highest standard deviations (0.189).

\begin{table*}[!htbp]
\centering
\caption{Performance in stress detection on CLAS dataset. P-values are calculated by ANOVA (post-hoc: Tukey)}
\label{table: clas}
\begin{tabular}{l|cccccc}
\toprule
      & Baseline & DANN & MMD & Triplet & MMD+Triplet & Person-Specific\\
    \midrule
    Accuracy & 64.3\% & 64.5\%& 66.5\% & 70.1\% & 70.6\% &86.8\%\\
     \midrule
     SD&0.012 &0.010& 0.014 & 0.011 & 0.010 & 0.189\\
     \midrule
      P-value $<$ 0.01& - 
      &$\times$
      &$\checkmark $ &$\checkmark $ & $\checkmark $ &
      $\checkmark $\\
     \bottomrule
\end{tabular}
\end{table*}

\subsection{Sleep Apnea Detection using ECG with Apnea-ECG Dataset}
Table \ref{table: ecg} shows the prediction accuracy of apnea detection using supervised learning models. The baseline model obtained a prediction accuracy of 75.2\% with a standard deviation of 0.014. The accuracy of the domain classification loss-based model showed a slight numerical increase but no statistically significant difference. The accuracy of the remaining three PiRL frameworks reached over 79\% and they were all statistically significantly higher than the baseline results with smaller standard deviations. The best framework for detecting apnea was the combination of MMD and triplet loss since it achieved the highest prediction accuracy. 
 The prediction accuracy of the person-specific model 
was highest which exceeded 95\% and statistically higher than the baseline. 

\begin{table*}[!htbp]
\centering
\caption{Performance in sleep apnea detection on Apena-ECG dataset. P-values are calculated by ANOVA (post-hoc: Tukey)}
\label{table: ecg}
\begin{tabular}{l|cccccc}
\toprule
      & Baseline & DANN & MMD & Triplet & MMD+Triplet & Person-Specific\\
    \midrule
    Accuracy & 75.2\% & 75.7\%
    & 79.5\% & 79.1\% & 79.9\% & 95.7\%\\
     \midrule
     SD& 0.014 & 0.013 & 0.010 & 0.013 & 0.009 & 0.011\\
     \midrule
     P-value $<$ 0.01& - 
      &$\times$
      &$\checkmark $ &$\checkmark $ 
      & $\checkmark $
      & $\checkmark $\\
     \bottomrule
\end{tabular}
\end{table*}

\subsection{Embedding Evaluation}

The effects of our proposed PiRL framework on latent space similarities among participants are shown in this section. As full subject ID information on the testing set was missing in the apnea dataset, we evaluated the latent space similarities only using the CLAS dataset. Table \ref{table: distance} illustrates the latent space similarities of invariant representations by cosine and manhattan distance among pair-wise participants in the baseline and MMD model to further explore the influence of MMD loss. Although the cosine distance of the MMD model (mean: 0.9981) was numerically less than the baseline model (mean: 1.0075), there was no statistical difference under the ANOVA test regarding the baseline as the reference group.  Meanwhile, the average value of the pair-wise manhattan distance is 380.3 and 306.9 for the baseline and MMD models respectively. The MMD model showed statistically shorter distance than the baseline model with the p-value equals to 2.2$\cdot$$10^{-16}$ under the ANOVA test. 
 \begin{table}[H]
\centering
\caption{Cosine and Manhattan Distance in the Baseline and MMD model on the CLAS dataset by ANOVA}
\label{table: distance}
\begin{tabular}{l|ccc}
\toprule
  & Baseline & MMD & P-value $<$ 0.01 \\
    \midrule
    Cosine Distance & 1.0075 & 0.9981 & $\times$\\
    \midrule
    Manhattan Distance & 380.27 & 306.94 & $\checkmark$ (2.2$\cdot$$10^{-16}$) \\
     \bottomrule
\end{tabular}
\end{table} 
The embeddings in the latent space of each participant turned got closer to each other using our PiRL framework. Consequently, by quantifying the embedding similarities in the latent space among participants using the manhattan distance, the results successfully reveal that MMD loss is capable of generating more participant-invariant representations on the generic models and thus, improving the performance. The ineffectiveness of cosine distance in the evaluation might be due to the loss of the coordinate information in the high dimensional space. The cosine similarity, which is based on cosine distance, measures the angle between vectors, therefore, if two embeddings have a small angle but are far apart in the original high dimensional space, they would not be considered similar.


\section{Conclusions and Future Work}

In this work, we proposed PiRL, which utilizes MMD and triplet loss for learning participant-invariant representations, to improve the performance of generic health detection models. We evaluated the performance and effectiveness of our framework using two public datasets for mental and physical health. As preliminary results, we demonstrated that our proposed PiRL outperformed the baseline models and helped generic models achieve better performances. Performance improvement was not observed using DANN technique. In future work, we will investigate other approaches to further optimize and interpret the representations in health applications.

\section*{Acknowledgment}
This work is supported by NSF \#2047296 and \#1840167.

\end{document}